\documentclass{elsart}
\usepackage{epsfig}
\newcommand{\ar}{\arrowvert}

\usepackage{amsfonts}
\usepackage{amsmath}
\usepackage{amssymb}
\usepackage{graphicx}
\newcommand{\FP}{\mathcal J}
\newcommand{\bs}{\boldsymbol}
\newcommand{\be}{\begin{equation}}
\newcommand{\ee}{\end{equation}}
\newcommand{\ba}{\begin{eqnarray}}
\newcommand{\ea}{\end{eqnarray}}

\usepackage{epsfig}
\usepackage{graphics}
\usepackage{latexsym}
\usepackage{amsmath}
\usepackage{amssymb}
\usepackage{rotating}
\usepackage{subfigure}

\usepackage{graphics}
\usepackage{latexsym}
\usepackage{rotating}
\usepackage{subfigure}

\begin{document}
\begin{frontmatter}
\hyphenation{english}
\title{Coulomb gauge model for hidden charm tetraquarks}
\author{W. Xie$^{a},$}
\author{L. Q. Mo$^{a}$,}
\author{Ping Wang$^{ab}$,}
\author{Stephen R. Cotanch$^{c}$}
\address{$^a$Institute of High Energy Physics, CAS, P. O. Box
918(4), Beijing 100049, China}
\address{$^b$Theoretical Physics Center for Science Facilities,
CAS, Beijing 100049, China}
\address{$ ^c$Department of Physics, North Carolina State
University, Raleigh, NC 27695-8202, USA}
\date{\today}
\maketitle
\begin{abstract}
The spectrum of tetraquark states with hidden charm is studied within an effective  Coulomb gauge Hamiltonian approach. Of the four independent color schemes, two are investigated, the $(q\bar{c})_1(c\bar{q})_1$ singlet-singlet  (molecule) and the $(qc)_3(\bar{q}\bar{c})_3$ triplet-triplet (diquark),  for selected $J^{PC}$ states using a variational method. The predicted masses of triplet-triplet tetraquarks are  roughly a GeV heavier than the singlet-singlet states. There is also an interesting flavor dependence with  $(q\bar{q})_1(c\bar{c}_1)$ states about half a GeV lighter than $(q\bar{c})_1(\bar{q}c)_1$.    The lightest  $1^{++}$ and $1^{--}$ predictions are in  agreement with the observed   $X(3872)$ and $Y(4008)$ masses suggesting they are molecules with   $\omega J/ \psi$ and $\eta h_c$, rather than $D^* \bar D^*$ and  $D\bar D$, type structure, respectively. Similarly, the  lightest isovector $1^{++}$ molecule , having a $\rho  J/ \psi$ flavor composition,  has mass near the recently observed charged $Z_c(3900)$ value.  These flavor configurations are consistent with observed $X$, $Y$ and $Z_c$ decays to $\pi \pi J/ \psi$.

\end{abstract}


\scriptsize{
PACS: 12.39.Mk; 12.39.Pn; 12.39.Ki;  12.40.Yx; 03.70.+k; 11.10.-z; 11.10.Gh


{\it{Keywords}}:  Exotic mesons, hidden charm, tetraquarks, QCD Coulomb gauge, effective Hamiltonian}
\end{frontmatter}





\section{Introduction}

\hspace*{\parindent}
With the recent observed  Higgs boson candidate at the LHC, the weak sector of the standard model may  be approaching closure.  However the strong interaction component is far from complete, especially  hadronic structure where intense searches are under way for  unconventional (non $q\bar q$ or $qqq$) states.
Indeed QCD allows for other  color singlet combinations, such as glueballs, hybrid mesons, tetraquarks, pentaquarks and dibaryons.  As early as 1977  $qq\bar{q}\bar{q}$ tetraquark states were proposed  \cite{Jaffe:1976ig} and today there are now  good 
candidates, the $\pi_1(1400)$ and $\pi_1(1600)$ \cite{Adams:1998ff,Chung:1999we},  listed in the PDG meson summary table \cite{PDG}. These states have unconventional quantum numbers $J^{PC} = 1^{-+}$ which are not possible for a $q\bar q$ system and have been reasonably described
\cite{General:2007bk} as $(q\bar{q})_1(q\bar{q})_1$ color singlet-singlet  states referred to as molecules. 
 In the heavy quark sector no states with unconventional quantum numbers have yet to be confirmed.  However, there are hidden charm states, some also listed in the PDG summary, which can not be described as $c\bar{c}$ mesons. These $X$, $Y$ and $Z$ particles are charmonium-like  but are not consistent with any   $c\bar{c}$ spectrum predictions. The lightest is $X(3872)$ which was discovered by Belle in 2003 \cite{Choi:2003ue} and has a narrow peak near 3872 $MeV$ in the $\pi^{+}\pi^{-}J/\psi$ invariant mass distribution from  $B^{-}\rightarrow K^{-}\pi^{+}\pi^{-}J/\psi$ decay. This discovery was confirmed by BaBar in the same decay process \cite{Aubert:2004zr}. Subsequently, additional $X$, $Y$ and $Z$ resonances were found, most recently
 \cite{Ablikim:2013} the first charged charmonium-like structure $Z_c(3900)$ which requires at least four quarks to have a non-zero electric charge. States with confirmed $J^{PC}$ are listed in Table \ref{table:XYZ}.
\begin{table}[htbp]
\centering
\caption{Properties of  $X$, $Y$ and $Z$ mesons with established $J^{PC}$.} \label{table:XYZ}
\begin{tabular}{|c|c|c|c|c|c|}
\hline
State    & Mass $(MeV)$     &$\Gamma(MeV)$&$J^{PC}$&   Decay   & Production $B\to$  \\
\hline
$X(3872)$&$3871.4\pm0.6$&$<2.3$&$1^{++} $&$\pi^{+}\pi^{-}J/\psi$&$ KX(3872)$\\
  &&&or $2^{-+}$&$\gamma J/\psi$& $p \bar{p}$\\
\hline
$Z(3930)$&$3929\pm5$&$29\pm10$&$2^{++}$&$D\bar{D}$&$\gamma\gamma$\\
\hline
$Y(4008)$&$4008^{+82}_{-49}$ &$226^{+97}_{-80}$&$1^{--}$&$\pi^{+}\pi^{-}J/\psi$&$e^{+}e^{-}$ \\
\hline
$Y(4260)$&$4260\pm12$&$83\pm22$&$1^{--}$&$\pi^{+}\pi^{-}J/\psi$&$e^{+}e^{-}$\\
\hline
$Y(4360)$&$4361\pm13$&$74\pm18$&$1^{--}$&$\pi^{+}\pi^{-}\psi^{\prime}$&$e^{+}e^{-}$\\
\hline
$Y(4660)$&$4664\pm12$&$48\pm15$&$1^{--}$&$\pi^{+}\pi^{-}\psi^{\prime}$&$e^{+}e^{-}$\\
\hline
\end{tabular}
\end{table}

Theoretically, there have been many $X$, $Y$ and $Z$ investigations using a variety of  methods, such as perturbative NRQCD \cite{Bodwin:1994jh,Brambilla:1999xf}, lattice QCD \cite{Okamoto:2001jb,Dudek:2007wv}, effective field theory \cite{Braaten:2006sy,Braaten:2005jj,Canham:2009zq}, QCD sum rule \cite{Matheus:2006xi,Lee:2008uy} and  potential models \cite{Liu:2008tn,Cao:2012du,Li:2009zu,Bhaghyesh:2011zzb},  all reviewed in detail elsewhere  \cite{Godfrey:2008nc,Swanson:2006st}.
One plausible conjecture \cite{Swanson:2003tb,Close:2003sg,Voloshin:2003nt}
 is these  states  are tetraquarks  containing light, $q =u$ or $d$, and charm, $c$,  quarks \ in different color schemes.
  In this paper we investigate this further and consider two  color combinations, the somewhat conventional molecular $(q\bar{c})_1(c\bar{q})_1$ singlet-singlet  \cite{Swanson:2006st} and the more exotic
$(qc)_{\bar{3}}(\bar{q}\bar{c})_3$ triplet-triplet (diquark) \cite{Maiani:2004vq}. We reserve the term exotic for the latter since it entails quarks in intermediate color states that are not singlets. 

Our work is an extension of a previous light tetraquark study \cite{General:2007bk}  utilizing the  Coulomb gauge [CG] model, first implemented to predict a glueball spectrum \cite{Szczepaniak:1995cw,Cotanch:1998ph} that was in good agreement with lattice QCD data.   In this approach the exact QCD Hamiltonian in the Coulomb gauge is replaced with an effective field theoretical relativistic Hamiltonian. The bare parton (current quark and gluon)  field operators are dressed by  a Bardeen, Cooper and Schrieffer [BCS] rotation 
and variational ground state (vacuum) minimization. 
This generates a non-trivial vacuum in which chiral symmetry is dynamically broken and quark/gluon constituent masses and condensates emerge. The hadrons are represented as quasiparticle excitations using many-body techniques such as Tamm-Dancoff and random phase approximations. This method was subsequently applied to mesons \cite{LlanesEstrada:1999uh,LlanesEstrada:2001kr,LlanesEstrada:2004}, hybrids \cite{Cotanch:2001mc,LlanesEstrada:2000hj,General:2006ed,Cotanch} and light tetraquark states \cite{General:2007bk,Wang:2008mw}.

This paper is organized into five sections. Section \ref{sec:model} specifies the CG model which is  then applied to hidden charm tetraquark states in section \ref{sec:tetra}.  Numerical results are  presented and discussed in section \ref{sec:numerical} followed by a summary, section \ref{sec:conclusion}, detailing conclusions and  future work.

\section{The QCD Coulomb gauge  model}\label{sec:model}

The CG model provides a comprehensive, systematic approach to hadron structure. It is applicable to both quarks and gluons,  light and heavy meson and baryon ground and excited states for any flavor and exotic systems involving different combinations of quarks and gluons in various color schemes.  It also permits consistent Hamiltonian dynamical mixing between states of entirely different dressed partons while providing insight characteristic of a wave function picture. As further discussed below there are additional attractive theoretical features and it is significant to note that there are no free model parameters as the
 two   dynamical constants,  the string tension $\sigma$ and Coulomb interaction $\alpha_s$, are predetermined from the literature.

The exact QCD Hamiltonian in the Coulomb gauge  \cite{T-D-Lee} is
\begin{eqnarray}
H_{\rm QCD} &=& H_q + H_g +H_{qg} + H_{C}    \\
H_q &=& \int d{\bf x} \Psi^\dagger ({\bf x}) [ -i
{\boldsymbol\alpha}\cdot
{\boldsymbol\nabla}
+  \beta m] \Psi ({\bf x})   \\
H_g &=& \frac{1}{2} \int\!\! d {\bf x} \!\! \left[ \FP^{-1}{\bf
\Pi}^a({\bf x})\cdot \!\!  \FP {\bf
\Pi}^a({\bf x}) +{\bf B}^a({\bf x})\cdot{\bf B}^a({\bf x}) \right] \\
H_{qg} &=&  g \int d {\bf x} \; {\bf J}^a ({\bf x})
\cdot {\bf A}^a({\bf x}) \\
H_C &=& -\frac{g^2}{2} \int d{\bf x} d{\bf y} \rho^a ({\bf x})
\FP^{-1} K^{ab}( {\bf x},{\bf y}  ) \FP \rho^b ({\bf y})   \ ,
\label{model}
\end{eqnarray}
where $g$ is the QCD coupling constant, $\Psi$ is the quark field with current quark mass $m$, $A^a=({\bf A}^a, A_0^a)$ are the gluon fields satisfying the Coulomb (transverse) gauge condition, $\boldsymbol{\nabla}\cdot \bf A^a = 0$, $(a=1,2,...8)$, ${\bf\Pi}^{a} = -{\bf E}_{tr}^a$ are the conjugate momenta and
\begin{equation}
 {\bf E}_{tr}^{a} = -\dot{\mathbf{A}}^a + g(1-\nabla^{-2} \boldsymbol{\nabla}\boldsymbol{\nabla}\cdot)f^{abc}A_0^b{\bf A}^c
\end{equation}
\begin{equation}
 {\bf B}^{a} = \boldsymbol{\nabla} \times {\bf A}^a + \frac12 g f^{abc}{\bf A}^b \times {\bf A}^c,
\end{equation}
are the non-abelian chromodynamic fields. The color densities $\rho^{a}({\bf x})$ and quark currents ${\bf J}^{a}({\bf x})$ are 
\begin{equation}
\rho^{a}({\bf x})= \Psi^{\dagger}({\bf x})T^{a}\Psi({\bf x})+f^{abc}{\bf A}^{b}({\bf x})\cdot {\bf \Pi}^{c}({\bf x})\\
\end{equation}
\begin{equation}
{\bf J}^{a}({\bf x}) = \Psi^{\dagger}({\bf x})\boldsymbol{\alpha}T^{a}\Psi({\bf x}),
\end{equation}
where $T^a = \frac{\lambda^a}{2}$ and $f^{abc}$ are the $SU(3)$
color matrices and structure constants, respectively. The
Faddeev-Popov determinant, $\FP = {\rm det}(\mathcal M)$, of the
matrix ${\mathcal M} = \boldsymbol{\nabla}\cdot
{\bf D}$, with covariant derivative ${\bf D}^{ab} =
\delta^{ab}\boldsymbol{\nabla}-gf^{abc} {\bf
A}^c$, is a measure of the gauge manifold curvature and the kernel
in Eq. (5) is given by $K^{ab}({\bf x}, {\bf y}) = \langle{\bf x},
a|{\mathcal M}^{-1} \nabla^2 {\mathcal M}^{-1}  |{\bf y}, b\rangle$.
The bare parton fields have the following normal mode expansions
(bare quark spinors $u, v$, helicity, $\lambda = \pm 1$, and color
vectors $\hat{\boldsymbol{\epsilon}}_{{\cal C }= 1,2,3}$)
\begin{eqnarray}
\label{colorfields1}
 \Psi(\bf{x})&=&\int \!\! \frac{d
    \bf{k}}{(2\pi)^3}[{u}_{\lambda}
({\bf k}) b_{\lambda \cal C}({\bf k})  +
{v}_{\lambda} (-{\bf k})
    d^\dag_{\lambda {\cal C}}(-{\bf k})]  e^{i {\bf k} \cdot \bf{x}} \hat{\boldsymbol{\epsilon}}_{\cal C}  \\
{\bf A}^a({\bs{x}}) &=&  \int \frac{d{\bs{k}}}{(2\pi)^3}
\frac{1}{\sqrt{2k}}[{\bf a}^a({\bs{k}}) + {\bf
a}^{a\dag}(-{\bs{k}})] e^{i{\bs{k}}\cdot {\bs {x}}}  \ \ \
\\
{\bf \Pi}^a({\bs{x}}) &=& \hspace{.15cm}-i \int
\!\!\frac{d{\bs{k}}}{(2\pi)^3} \sqrt{\frac{k}{2}} [{\bf
a}^a({\bs{k}})-{\bf a}^{a\dag}(-{\bs{k}})]e^{i{\bs{k}}\cdot
{\bs{x}}}  \!.
\end{eqnarray}

Our model's starting point is the Coulomb gauge QCD Hamiltonian.
We then make the following substitutions: 1)
replace the exact Coulomb kernel with a calculable confining potential; 2)
use the lowest order, unit value for the Faddeev-Popov determinant.
Therefore, $H_C$ in the CG model Hamiltonian becomes
\begin{eqnarray}
H_C^{\rm CG} &=& -\frac{1}{2} \int d{\bf x} d{\bf y} \rho^a ({\bf
x}) \hat{V}(\ar {\bf x}-{\bf y} \ar ) \rho^a ({\bf y}) .
 \label{model}
\end{eqnarray}
The confining and leading canonical interaction is represented by a Cornell type potential $\hat{V}(r)=-\alpha_s/r+\sigma r$. Previous studies with this interaction were in good agreement with both lattice glueball  masses  \cite{Szczepaniak:1995cw}
and the observed meson spectrum \cite{LlanesEstrada:2001kr}.
Performing a fourier transform, the potential in  momentum space is
\begin{equation}
 V(|{\bf k}|)=-\frac{4\pi\alpha_s}{k^2}-\frac{8\pi\sigma}{k^4}.
\end{equation}

Next, hadron states are expressed as BCS vacuum $|\Omega\rangle$ excitations involving dressed quark (antiquark) Fock operators $B^{\dagger}_{\lambda \cal C}$,
 $D_{\lambda\cal C}$ 
related to the bare operators $b_{\lambda\cal C}$, $d^{\dag}_{\lambda\cal C}$ by the rotation
\begin{eqnarray}
B_{\lambda \cal C}(\bf k)&=&\cos\frac{\theta(k)}{2}b_{\lambda\cal C}({\bf k})-\lambda \sin\frac{\theta(k)}{2}d^{\dag}_{\lambda\cal C}(-{\bf k})\\
D_{\lambda\cal C}(-{\bf k})&=&\cos\frac{\theta(k)}{2}d_{\lambda\cal C}(-{\bf k})+\lambda \sin\frac{\theta(k)}{2}b^{\dag}_{\lambda\cal C}({\bf k}),
\end{eqnarray}
where $\theta(k)$ is the BCS angle.  Correspondingly, the dressed Dirac spinors are
\begin{eqnarray}
 \cal U_{\lambda}({\bf k})&=&\frac{1}{\sqrt 2}
 \left(
  \begin{array}{c}
    \sqrt{1+\sin\phi(k)}\chi_{\lambda}\\
    \sqrt{1-\sin\phi(k)}\boldsymbol{\sigma}\cdot\hat{\bf k}\chi_{\lambda}
  \end{array}
  \right)\\
\cal V_{\lambda}({-\bf k})&=&\frac{1}{\sqrt 2}
\left(
  \begin{array}{c}
    -\sqrt{1-\sin\phi(k)}\boldsymbol{\sigma}\cdot\hat{\bf k}\chi_{\lambda}\\
    \sqrt{1+\sin\phi(k)}\chi_{\lambda}
  \end{array}
  \right) ,
\end{eqnarray}
where $\phi (k)$ is the  gap angle from the  gap equation
that minimizes the energy of the BCS vacuum \cite{General:2006ed} and is related to $\theta(k)$ by $tan(\phi(k) - \theta(k)) = m / k$.

The four model parameters have all been predetermined  \cite{LlanesEstrada:2004}.  The light and charm bare quark masses are  $5\; MeV$ and $1350 \; MeV$, respectively,  while the two dynamic constants are $\sigma=0.18 \; GeV^2$ and $\alpha_s=\frac{g^2}{4\pi}=0.4$. 
This few parameter, constrained model has successfully described  meson, hybrid, glueball and light tetraquark systems
yielding results agreeing with  experimental and lattice data. 

\section{Application to hidden charm tetraquark states}\label{sec:tetra}
\begin{figure}[t]
    \vspace{-3cm}
    \includegraphics[width=1.0\textwidth]{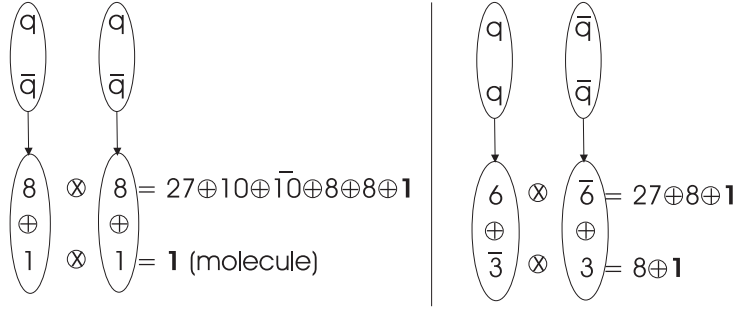}
 \vspace{-4cm}
\caption{Four tetraquark color  schemes. One is a  singlet-singlet molecule while the other three are more exotic atoms (octet and two diquark schemes).}
\label{fig:colorwf}
\end{figure}

For this heavier tetraquark system, the quark (antiquark) $cm$ momenta are ${\bf{k}}_{i = 1, 2, 3, 4}$ 
and the following wave function ansatz is adopted
\begin{eqnarray}
    |\Psi^{JPC}\rangle=\int d{\bf{k}}_1 d{\bf{k}}_2 d{\bf{k}}_3 d{\bf{k}}_4  \delta({\bf{k}}_1 + {\bf{k}}_2 + {\bf{k}}_3 +{ \bf{k}}_4)
    \Phi^{JPC}_{\lambda_1 \lambda_2
    \lambda_3 \lambda_4}({\bf{k}}_i) \notag \\
   \times R^{{\cal C}_1{\cal C}_2}_{{\cal C}_3{\cal C}_4}
    B^{\dag}_{\lambda_1{\cal C}_1}({\bf{k}}_1)
    B^{\dag}_{\lambda_2{\cal C}_2}({\bf{k}}_2)
    D^{\dag}_{\lambda_3{\cal C}_3}({\bf{k}}_3)
    D^{\dag}_{\lambda_4{\cal C}_4}({\bf{k}}_4)|\Omega \rangle  .
 \end{eqnarray}
The expression for the matrix $R^{{\cal C}_1{\cal C}_2}_{{\cal C}_3{\cal C}_4}$ depends on the specific color scheme selected.
As depicted in  Fig. \ref{fig:colorwf},
the $SU_c(3)$ color algebra for four quarks produces 81 color states, ${3}\otimes \bar{3}\otimes{3}\otimes \bar{3} = {27} \oplus{10} \oplus { \bar {10} }\oplus 8\oplus 8 \oplus 8 \oplus 8 \oplus 1 \oplus1$ of which two are overall color singlets that can be obtained
in four different schemes (only two are independent) depending on the intermediate  coupling of two quarks:
singlet-singlet (molecule), octet-octet and two diquark schemes triplet-triplet and sextet-sextet.
In this article we focus on singlet-singlet (molecule) and triplet-triplet (diquark) schemes, which provide the lightest states.
 
For the diquark configuration, the  wave function  is
\begin{align}
\!\! \! \!\! \! \Phi^{JPC}_{\lambda_1 \lambda_2\lambda_3 \lambda_4}({\bf{k}}_i)= F^{JPC}({\bf{k}}_i)\langle \frac{1}{2}\frac{1}{2}\lambda_1\lambda_2|S_A \lambda_{A}\rangle\langle \frac{1}{2}\frac{1}{2}\lambda_3\lambda_4|S_B\lambda_{B}\rangle\langle S_A S_B \lambda_{A} \lambda_{B}|S \lambda\rangle\notag\\
\langle L_A L_B m_{A} m_{B}|l m\rangle\langle l L_I m m_{I}|L M\rangle\notag \langle S L \lambda M|J M_J\rangle Y_{L_A}^{m_A}({\hat{\bf k}}_A)Y_{L_B}^{m_B}({\hat{\bf k}}_B)Y_{L_I}^{m_I}({\hat{\bf k}}_I),
\end{align}
where $Y_L^m({\hat{\bf k}})$ is the spherical harmonic function, while $S_{A(B)}$, $L_{A(B)}$ and $L_I$ are the $A$($B$)  diquark total spin, orbit angular momentum 
and orbit angular momentum between the two diquarks, respectively.
The radial  wave function, $F^{JPC}({\bf{k}}_i)$,
is chosen to be a Gaussian, exp$(-\frac{{k}_A^2}{\alpha_A^2}-\frac{{ k}_B^2}{\alpha_B^2}-\frac{{ k}_I^2}{\alpha_I^2})$,
where ${\bf k}_A$, ${\bf k}_B$ and ${\bf k}_I$ are respectively
${\bf k}_A=\frac{{\bf k}_1-{\bf k}_2}{2}$, ${\bf k}_B=\frac{{\bf k}_3-{\bf k}_4}{2}$,
 ${\bf k}_I=\frac{{\bf k}_1+{\bf k}_2}{2}-\frac{{\bf k}_3+{\bf k}_4}{2}$ and 
$\alpha_A$, $\alpha_B$ and $\alpha_I$ are the variational parameters.
The molecular wavefunction is similar and now $A$ and $B$ denote the $(q \bar q)_1$ and
$(c \bar c)_1$ quantities, respectively. The $(q \bar c)_1({\bar q} c)_1$ molecule is symmetric in $A$ and $B$. 

\begin{figure}[t]
\centering
\vspace{-8.5cm}
\includegraphics[width=10cm]{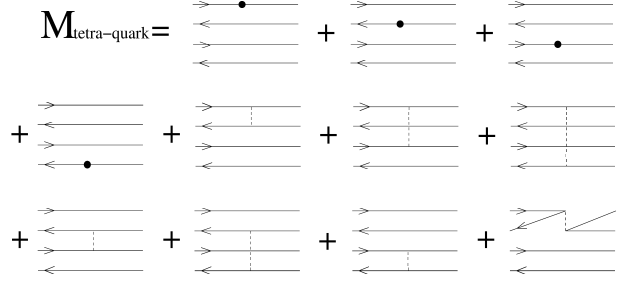}
\caption{Equal-time diagrams for tetraquarks. The first four diagrams are for the kinetic and self energy. The last diagram is for the
annihilation term and the others are for the $qq$, $q \bar q$ and $\bar q \bar q$ interactions.}
\label{pic:diag0}
\end{figure}

The variational mass is then given by
\begin{eqnarray}
M_{J^{PC}}&=&\frac{\langle\Psi^{JPC}|H^{CG}|\Psi^{JPC}\rangle}{\langle\Psi^{JPC}|\Psi^{JPC}\rangle} = M_{self} + M_{qq}
 + M_{q\bar q} + M_{\bar q \bar q} + M_{annih} \;  ,
\end{eqnarray}
where the subscripts indicate the contribution from the single particle quark self energy and two particle $qq$, $q \bar q$, $\bar q \bar q$ scattering and $q \bar q$ annihilation, respectively.  All contributions to the tetraquark mass are diagramed in Fig.~\ref{pic:diag0}.

Using the above specified Hamiltonian and wave function, the self energy term can be  reduced to
\begin{eqnarray}
 \!\!\! \! M_{self}&=& \sum_{i = 1} ^4\int\frac{d{\bf k}_1}{(2\pi)^3}\frac{d{\bf k}_2}{(2\pi)^3}\frac{d{\bf k}_3}{(2\pi)^3}\frac{d{\bf k}_4}{(2\pi)^3}
    \Phi^{\dag JPC}_{\lambda_1\lambda_2\lambda_3\lambda_4}({\bf k}_1,{\bf k}_2,{\bf k}_3,{\bf k}_4) \notag \\
    && \times 
      E_i ({\bf k}_i)   \Phi^{JPC}_{\lambda_1\lambda_2\lambda_3\lambda_4}({\bf k}_1,{\bf k}_2,{\bf k}_3,{\bf k}_4) ,
\end{eqnarray}
where the single particle energy is given by
\begin{eqnarray}
\!\!\! \! E_i({\bf k}_i) &=&m_i sin \,  \theta({\bf k}_i) + k_i cos \,  \theta({\bf k}_i)   -\frac{2}{3}\int\frac{d{\bf q}}{(2\pi)^3} V(|{\bf q}|) [sin \,  \theta({\bf k}_i) \sin  \,  \theta({ \bf q}) \notag \\ && + \;  cos \,  \theta({\bf k}_i) cos  \,  \theta({ \bf q}) 
  \hat{\bf q}\cdot \hat{\bf k}_i]   .
\end{eqnarray}
Similarly, the two particle contributions can be evaluated yielding
\begin{eqnarray}
 \!\!\! \! M_{qq}&=&-\frac{2}{3}\int\frac{d{\bf k}_A}{(2\pi)^3}\frac{d{\bf k}_B}{(2\pi)^3}\frac{d{\bf k}_I}{(2\pi)^3}\frac{d{\bf q}}{(2\pi)^3}
     \mathcal U_{\lambda_1^{\prime}}^{\dag}({\bf k}_{1}^{\prime}) \mathcal U_{\lambda_1}({\bf k}_{1})
          \mathcal U_{\lambda_2^{\prime}}^{\dag}({\bf k}_{2}^{\prime}) \mathcal U_{\lambda_2}({\bf k}_{2}) \notag \\
         &&\times V(|{\bf q}|)\Phi^{\dag JPC}_{\lambda_1\lambda_2\lambda_3\lambda_4}({\bf k}_1,{\bf k}_2,{\bf k}_3,{\bf k}_4)
         \Phi^{JPC}_{\lambda_1^{\prime}\lambda_2^{\prime}\lambda_3^{\prime}\lambda_4^{\prime}}({\bf k}_1^{\prime},{\bf k}_2^{\prime},{\bf k}_3^{\prime},{\bf k}_4^{\prime}) ,
\end{eqnarray}
\begin{eqnarray}
  \!\!\! \!M_{q\bar q}&=&\frac{1}{3}\int\frac{d{\bf k}_A}{(2\pi)^3}\frac{d{\bf k}_B}{(2\pi)^3}\frac{d{\bf k}_I}{(2\pi)^3}\frac{d{\bf q}}{(2\pi)^3}
           \mathcal U_{\lambda_1^{\prime}}^{\dag}({\bf k}_{1}^{\prime}) \mathcal U_{\lambda_1}({\bf k}_{1})
         \mathcal V_{\lambda_3}^{\dag}({\bf k}_{3}) \mathcal V_{\lambda_3^{\prime}}({\bf k}_{3}^{\prime})  \notag \\
         &&\times V(|{\bf q}|)\Phi^{\dag JPC}_{\lambda_1\lambda_2\lambda_3\lambda_4}({\bf k}_1,{\bf k}_2,{\bf k}_3, {\bf k}_4)
         \Phi^{JPC}_{\lambda_1^{\prime}\lambda_2^{\prime}\lambda_3^{\prime}\lambda_4^{\prime}}({\bf k}_1^{\prime},{\bf k}_2^{\prime},{\bf k}_3^{\prime},{\bf k}_4^{\prime}),
\end{eqnarray}
\begin{eqnarray}
 \!\!\! \! M_{\bar q \bar q}&=&-\frac{2}{3}\int\frac{d{\bf k}_A}{(2\pi)^3}\frac{d{\bf k}_B}{(2\pi)^3}\frac{d{\bf k}_I}{(2\pi)^3}\frac{d{\bf q}}{(2\pi)^3}
     \mathcal V_{\lambda_1^{\prime}}^{\dag}({\bf k}_{1}^{\prime}) \mathcal V_{\lambda_1}({\bf k}_{1})
          \mathcal V_{\lambda_2^{\prime}}^{\dag}({\bf k}_{2}^{\prime}) \mathcal V_{\lambda_2}({\bf k}_{2}) \notag \\
         &&\times V(|{\bf q}|)\Phi^{\dag JPC}_{\lambda_1\lambda_2\lambda_3\lambda_4}({\bf k}_1,{\bf k}_2,{\bf k}_3,{\bf k}_4)
         \Phi^{JPC}_{\lambda_1^{\prime}\lambda_2^{\prime}\lambda_3^{\prime}\lambda_4^{\prime}}({\bf k}_1^{\prime},{\bf k}_2^{\prime},{\bf k}_3^{\prime},{\bf k}_4^{\prime}) ,
\end{eqnarray}
where ${\bf q}$ is the momentum transfer ${\bf k}_1^{\prime} - {\bf k}_1$.
Finally, the annihilation term is 
\begin{eqnarray}
\label{annih}
 \!\!\! \! M_{annih}&=&\frac{1}{3}\int\frac{d{\bf k}_A}{(2\pi)^3}\frac{d{\bf k}_B}{(2\pi)^3}\frac{d{\bf k}_I}{(2\pi)^3}\frac{d{\bf q}}{(2\pi)^3}
         \times \mathcal V_{\lambda_4}^{\dag}({\bf k}_{4}) \mathcal U_{\lambda_1}({\bf k}_{1})
         \mathcal U_{\lambda_1^{\prime}}^{\dag}({\bf k}_{1}^{\prime}) \mathcal V_{\lambda_4^{\prime}}({\bf k}_{4}^{\prime})  \notag \\
         &&\times V(|{\bf q}|)\Phi^{\dag JPC}_{\lambda_1\lambda_2\lambda_3\lambda_4}({\bf k}_1,{\bf k}_2,{\bf k}_3,{\bf k}_4)
         \Phi^{JPC}_{\lambda_1^{\prime}\lambda_2^{\prime}\lambda_3^{\prime}\lambda_4^{\prime}}({\bf k}_1^{\prime},{\bf k}_2^{\prime},{\bf k}_3^{\prime},{\bf k}_4^{\prime}) ,
\end{eqnarray}
where now {\bf q} is the annihilation momentum ${\bf k}_1+{\bf k}_4$. For the molecular case, each $q \bar q$ pair forms
a subgroup and the corresponding wave function and  mass formula are similar. The  12-dimension integrals are performed numerically using the Monte-Carlo method.

\section{Numerical Results}\label{sec:numerical}

\begin{table}[b]
\centering
\caption{Predicted spectrum for tetraquarks (singlet-singlet and triplet-triplet) and hybrid mesons in $MeV$. Unless explicitly stated,
all $L_i$ and $S_i$ are zero.}\label{table:data}
\vspace{.25cm}
\begin{tabular}{|||c||c|c||c|c|c||||c|c||}
\hline
$L_i,S_i$   & $J^{PC}$     & $(q\bar q)_1(c\bar c)_1$   & $J^{PC}$ &  $(q\bar c)_1(c\bar q)_1$ & $(qc)_{\bar 3}(\bar c\bar q)_3$
& $J^{PC}$ & $(c \bar c)_8 g $  \\
\hline
\hline
$L_i,S_i=0$  & $0^{++}$   &  3540 & $0^{++}$ &  4129 &  4428 & $0^{++}$ & 3945   \\
\hline
 $ L_I=1$  & $1^{-+}$   &  3540 & $1^{--}$ &  4129 &  4741 & $1^{-+}$  & 4020 \\
\hline
  $L_A=1$  & $1^{--}$   &  4186 & $1^{-+}$ &  4563  &  4826 & $1^{-+}$  & 4155 \\
\hline
 $L_B=1$  & $1^{--}$   &  4026 & $1^{-+}$ &  4565 &  4825 & $1^{-+}$  & 4565  \\
\hline
 $S_A=1$  & $1^{+-}$   &  3781 & $1^{++}$ &  4137 &  4430 & $1^{++}$  & 4100   \\
\hline
 $S_B=1$  & $1^{+-}$   &  3538 & $1^{++}$  &  4137 &  4430 & $1^{+-}$  & 3830 \\
\hline
$S_A, S_B=1$  & $0^{++}$   &  3783 & $0^{++}$ &  4144 &  4405 & $0^{--}$  & 4020  \\
\hline
 $S_A, S_B=1$  & $1^{++}$   &  3783 & $1^{++}$ &  4144 &  4419 & $3^{-+}$  & 4615  \\
\hline
 $S_A, S_B=1$  & $2^{++}$   &  3783 & $2^{++}$  &  4144 &  4444 & $2^{++}$  & 3965  \\
\hline
\end{tabular}
\end{table}

We calculated all  possible states with $S_i=0,1$ and/or $L_i=0,1$ and the numerical results for corresponding $J^{PC}$ are listed in Table \ref{table:data}. For model completeness and further comparative insight, we also list our $c \bar c g$ hybrid predictions from an earlier publication \cite{General:2006ed}.
It is clear that the diquark composition is heavier than the corresponding molecular state.
Even more significant is the sensitivity to arrangement of flavor as $(q\bar{q})_1(c\bar{c}_1)$ states are much lighter than
$(q\bar{c})_1(\bar{q}c)_1$ for the same $L_i $, $S_i$ configuration. 
 The CG model therefore predicts that the lightest tetraquark states
are molecules having quarks with the same flavor in a color singlet (e.g.  a $\eta \eta_c$ or $\omega J/\psi$ type  flavor structure depending upon spin).

For the $s$ wave triplet-triplet tetraquark states, the mass changes little with different spin. This is because there is a heavy charm quark 
in each color triplet  and the spin splitting decreases with heavier quark mass.
For orbit angular momentum 1, the  triplet-triplet tetraquark mass increases about 300 to 400 $MeV$ and
all  $p$ wave tetraquarks have mass  above 4400 $MeV$.

Similarly for the $(q\bar{c})_1(c\bar{q})_1$ states, the spin splitting is very small due to the large charm quark mass in each color singlet,
and again if either $L_A$ or $L_B$ changes from 0 to 1 the mass increases, now
about 430 $MeV$. For the $(q\bar{q})_1(c\bar{c})_1$ flavor arrangement, the mass does not depend on the $c\bar{c}$ spin   but
increases about 240 $MeV$ if the spin of light quark subgroup changes from 0 to 1. The mass increase due to larger orbit angular momentum 
is also flavor sensitive, since for $q\bar{q}$ $p$ waves  it is about 650 $MeV$, while  for $c\bar{c}$ $p$ waves it
is 490 $MeV$.

\begin{figure}[b]
\centering
\includegraphics[width=8.6cm]{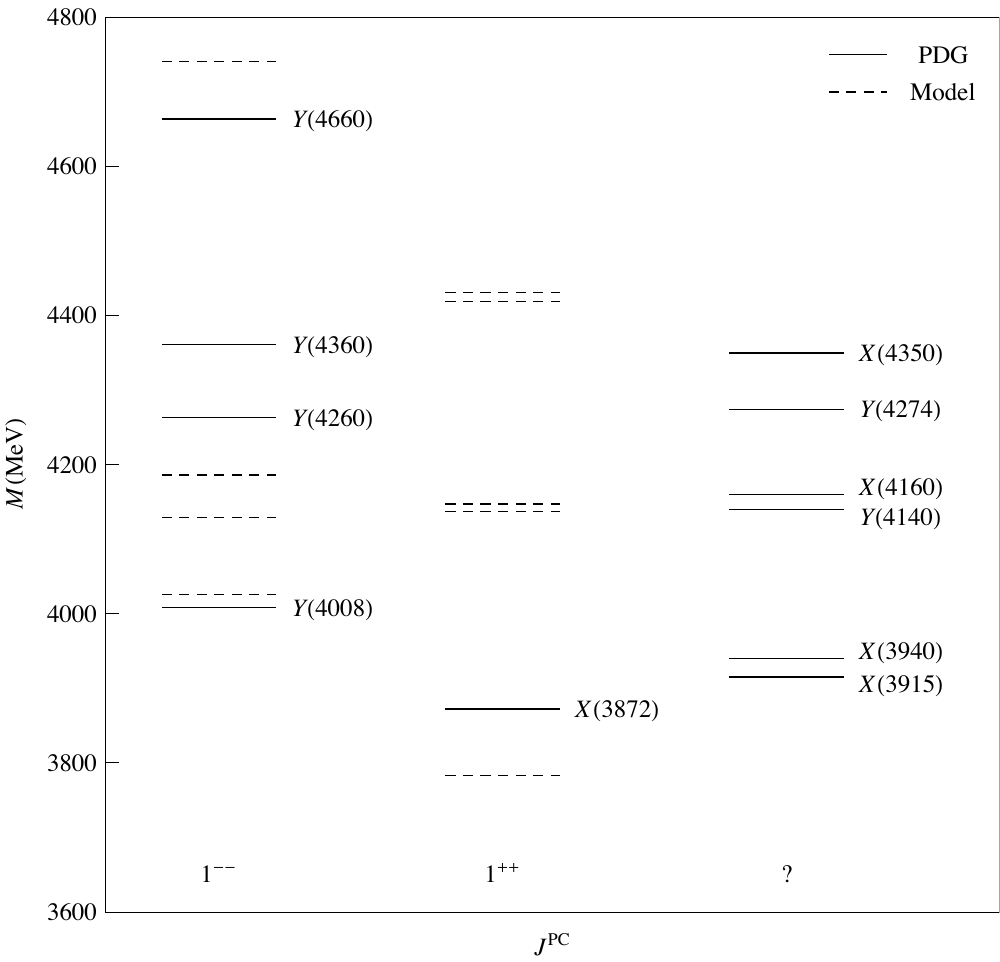}
\caption{Theoretical results (dashed lines) for $1^{++}$ and $1^{--}$ tetraquark states compared to data (solid lines).}\label{pic:data}
\end{figure}

In Fig. \ref{pic:data}, we compare predicted $1^{++}$ and $1^{--}$ tetraquark masses to
observed $X$ and  $Y$ states. Only about half of the states have established $J^{PC}$ and to  aid
 determining quantum numbers it is noteworthy  that the observed $Y(4140)$, $X(4160)$,
$Y(4274)$ and $X(4350)$ are comparable to the heaviest model $1^{++}$ states.
If this identification proves correct, then the model predicts the $Y(4140)$ and  $X(4160)$ are molecular 
states  consisting of either one or two spin 1 clusters having a $D\bar D$ flavor composition while the
 $Y(4274)$ and $X(4350)$ would be exotic  triplet-triplet states with a diquark  composition. 
 
 Regarding the established $J^{PC}$ states, the predicted $(q\bar{q})_1(c\bar{c})_1$ $1^{++}$ mass   is 3783 $MeV$, about $2\%$ less than observed,  suggesting the
$X(3872)$ may be a molecular state with a $\omega J/ \psi$ flavor composition.
Since the annihilation mass contribution, Eq. (\ref{annih}), does not contribute in a $q\bar q$ color singlet or isospin 1 configuration, our model also predicts a degenerate isovector $1^{++}$ state,
now having a $\rho J/ \psi$ molecular type structure, with mass near the recently observed \cite{Ablikim:2013} charged $Z_c(3900)$.

  Four $1^{--}$ tetraquark states are predicted with  the lightest mass
at 4026 $MeV$ which is close to $Y(4008)$ discovered by Belle \cite{Yuan:2007sj}. The masses of the next two heavier molecular states are 4129 and 4186, which are
comparable with the unknown $J^{PC}$ $Y(4140)$ and $Y(4260)$. The states with masses 4026 $MeV$ and 4186 $MeV$   have  $\eta h_c$ and $h_1 \eta_c$ type structures, respectively,   while the state at 4129 $MeV$ is a $D \bar D$ type molecule. Because the $Y(4260)$ decays to $\pi \pi J/\psi$ rather than $D \bar D$, it is tempting to identify it with the model $Y(4186)$ especially if the  $Y(4140)$ has $J^{PC} = 1^{++}$, in contrast to $1^{++}$ discussed above. If so a  mixing analysis may be necessary since these two model states are close in mass.  The calculated mass of the triplet-triplet $1^{--}$ state is 4741 $MeV$ which is also near to the  $Y(4660)$ mass. This state would  be an exotic diquark  if this assignment prevails. Lastly, for the observed $2^{++}$ $Z(3930)$ listed in Table \ref{table:XYZ}, the model tetraquark states are either above or below this mass by over 150 $MeV$, however the predicted 
$c \bar c g$ hybrid mass of 3965 $MeV$ (see Table \ref{table:data})  agrees quite well. Similarly
the $0^{++}$ hybrid at 3945 $MeV$ is very close to the $X(3915)$. This state does not have established quantum numbers other than the $G = + 1$ parity which is also the same as our hybrid prediction.  The hybrid results are quite firm since they include both hyperfine and the 
non-linear (non-Abelian) component of the color magnetic fields (Eq. (7)).

Finally, it is interesting  that the decay channels summarized in Table 1 are consistent with the Coulomb gauge model flavor structure predictions for those observed states. Specifically, for the predicted molecular state assignments involving the  $X(3872)$, $Z_c(3900)$, $Y(4008)$ and $Y(4260)$ having a non $D \bar D$ or $D^* \bar D^*$ flavor structure,  their  decays would favor the observed $\pi \pi J/\psi$ over  channels with $D$ mesons. Related, the relative values of the $2 \pi$ widths,   small, 1.2 $MeV$, for the $X(3872)$  and over an order of magnitude larger, 46 $MeV$ \cite{Ablikim:2013}, for the $Z_c(3900)$,  are consistent with a $1^{++}$ assignment since 2$\pi$ decay is isospin violating for $I$ = 0 but allowed for $I = 1$. For the diquark model states, detailed decay calculations would be necessary to make firm predictions. Similarly, for the undetermined $J^{PC}$ states,
$Y(4140)$ and $X(4160)$, knowing their decays would be a helpful identifying signature.

\section{Summary}\label{sec:conclusion}

We have applied the Coulomb gauge model to  tetraquark systems with hidden charm having two different color compositions,  singlet-singlet and triplet-triplet. Generally,  triplet-triplet states are the heaviest for the same spin or orbit angular momentum. For color singlet-singlet states, we
further investigated the flavor groupings corresponding to $(q \bar q)_1(c \bar c)_1$ versus $(q {\bar c})_1({\bar q} c)_1$  type molecules (meson-meson states) and found that the same flavor color singlets produce a lower mass. Also, the mass from the spin part decreases rapidly with  increasing quark mass, while  the orbit angular momentum contribution decreases slowly with 
increasing quark mass.

Our key finding is  the  $X(3872)$ mass is close to the predicted $1^{++}$ $(q\bar{q})_1(c\bar{c})_1$ state suggesting it is a conventional $\omega J/\psi$ type molecule and not a more exotic diquark tetraquark state. Four $1^{--}$ states are predicted,
three  molecular  and one  triplet-triplet. Their masses are comparable with $Y(4008)$, $Y(4260)$, $Y(4360)$ and $Y(4660)$.
The masses of  other states with undetermined $J^{PC}$, most significantly the charged  $Z_c(3900)$, are close to the calculated $1^{++}$ results. Therefore, the $J^{PC}$ number of these states could be $1^{++}$.

Future work will address several issues. The spin dependence will be further studied now including the hyperfine interaction \cite{LlanesEstrada:2004}. Also, a dynamic mixing analysis will be performed involving $q \bar q$ and $qq\bar q\bar q$ states having various flavor distributions as well as mixing with glueball and hybrid meson exotic states. Finally, the CG model will be applied to  $b$ quark  systems to aid experimental searched at higher energies.

\section*{Acknowledgement} The authors are deeply grateful to Felipe Llanes-Estrada for insightful comments. This work is supported in part by DFG and NSFC (CRC 110)
and by National Natural Science Foundation of China (Grant No. 11035006).

\end{document}